\documentclass[prl,twocolumn,aps,amssymb,showpacs,superscriptaddress]{revtex4}
\usepackage{graphicx}
\usepackage{color}

\begin{document}

\title{Quasi-Eigenstate Evolution in Open Chaotic Billiards}

\author{Sang-Bum Lee}
\address{School of Physics and Astronomy, Seoul National University, Seoul
151-742, Korea}
\author{Juhee Yang}
\address{School of Physics and Astronomy, Seoul National University, Seoul 151-742, Korea}
\author{Songky Moon}
\address{School of Physics and Astronomy, Seoul National University, Seoul 151-742, Korea}
\author{Soo-Young Lee}
\address{School of Physics and Astronomy, Seoul National University, Seoul 151-742, Korea}
\author{Jeong-Bo Shim}
\address{Max Planck Institute for the Physics of Complex Systems, N\"{o}thnitzer Str. 38, Dresden, Germany}
\author{Sang Wook Kim}
\address{Department of Physics Education and Department of Physics, Pusan National University, Busan 609-735, Korea}
\author{Jai-Hyung Lee}
\address{School of Physics and Astronomy, Seoul National University, Seoul 151-742, Korea}
\author{Kyungwon An}
\email{kwan@phya.snu.ac.kr}
\address{School of Physics and Astronomy, Seoul National University, Seoul 151-742, Korea}

\date{\today}

\begin{abstract}
\noindent We experimentally studied evolution of quasi-eigenmodes
as classical dynamics undergoing a transition from being regular
to chaotic in open quantum billiards. In a deformation-variable
microcavity we traced all high-$Q$ cavity modes in a wide range of
frequency as the cavity deformation increased. By employing an
internal parameter we were able to obtain a mode-dynamics diagram
at a given deformation, showing avoided crossings between
different mode groups, and could directly observe
the coupling strengths induced by ray chaos among encountering
modes. We also show that the observed mode-dynamics
diagrams reflect the underlying classical ray dynamics in the phase
space.
\end{abstract}


\maketitle


Quantum manifestation in a classically chaotic system has become
an important issue in atomic, nano, mesoscopic physics, etc., due
to its fundamental importance in quantum mechanics and
applications to practical quantum/wave systems \cite{Q-chaos}.
Most of early works have focused on statistical analysis of
eigenvalues and eigenfunctions and comparison with the random
matrix theory, {\em e.g.}, the transition from Poisson to Wigner
distribution of level spacings during a transition to chaos,
providing an averaged view on mode dynamics \cite{Q-chaos}.
Experimental verifications of the statistics have been performed
mainly in {\em closed} microwave cavities \cite{stoeckmann}.
Dynamical tunneling or coupling between regular and
chaotic modes has recently been observed for a mixed phase space
specially tailored for this purpose \cite{backer}.

In {\em open} quantum systems, each quasi-eigenmode has a
linewidth, and thereby changes the mode dynamics significantly.
Trapped modes were observed showing high $Q$ even with increasing
coupling strength to open channels in microwave cavities
\cite{Persson}, and crossing and avoided crossing (AC) of cavity
modes were reported near an exceptional point formed by two
coupled microwave cavities \cite{Dembowiski}. We note, however,
that the previous experimental works in microwave cavities and
other systems neither realized an optimal system showing a
continuous chaotic transition from being regular to chaotic nor
provide observations direct enough to tell the variation of
statistics.

In this paper, we have experimentally observed, for the first time, the evolution of
quasi-eigenmode dynamics in a generic open
nonintegrable system when classical dynamics undergoes a
transition from being regular to fully chaotic. In a
dielectric deformation-variable chaotic optical
microcavity (COM) we traced all high-$Q$ cavity modes in a wide
range of frequency as the cavity deformation increases. By
introducing an additional parameter orthogonal to the cavity
deformation, we could explicitly observe  mode-mode
dynamics under the chaotic transition and measure
various mode-mode coupling constants which can be associated with
the underlying classical ray dynamics in phase space. We believe
our data would be a valuable asset for future formulation of a
currently-nonexisting semiclassical theory for
coupling strengths between modes in a mixed phase space.

Our experiment was performed in a two-dimensional
COM made of a liquid jet column\cite{noeckel} of
ethanol (refractive index $m$=1.361) doped with either Rhodamine B
dye at a concentration of $10^{-7}$mol/cm$^3$ or Rhodamine 6G dye
at a concentration of $10^{-9}$mol/cm$^3$, depending on the
wavelength region of interest. Its boundary is approximated by
$r(\phi)\simeq a(1+\eta\cos2\phi+\epsilon\eta^2\cos4\phi)$ in the
polar coordinates with $a\simeq$14.9$\pm0.1\mu$m and
$\epsilon=0.42\pm0.05$ \cite{Yang-RSI06, Moon-OE08}. The
deformation parameter $\eta$ can be continuously varied from 0\%
to 26\%. The size parameter defined as $mka$ with $k=2\pi/\lambda$
and wavelength $\lambda\sim$600 nm is about 200, thus comprising
the short wavelength limit. We measured cavity-modified
fluorescence (CMF) and/or lasing spectra by using
the method described in Refs.\ \cite{Lee02, Lee07PRA}.

\begin{figure}
\includegraphics[width=3.5in]{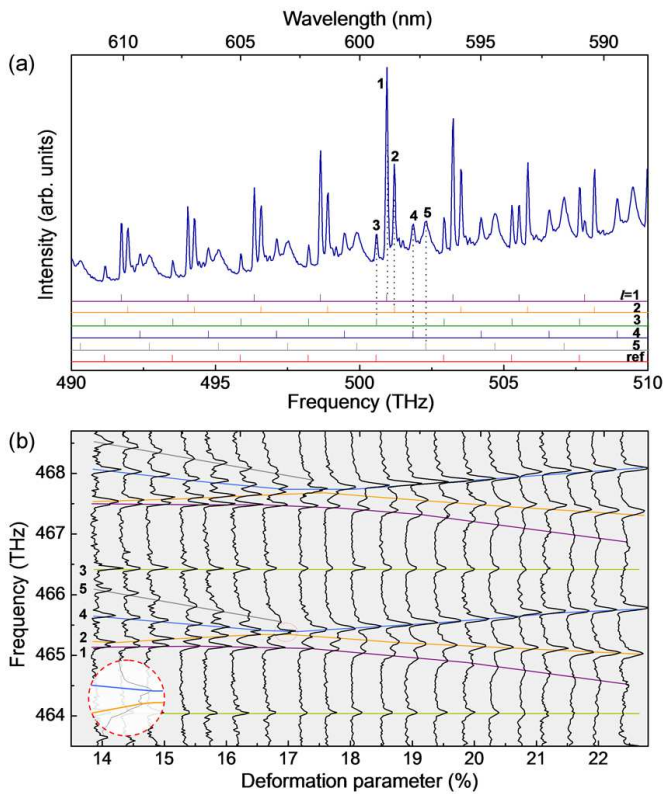}
\caption{(Color on line) (a) For a COM with $\eta$=18.7\%, five uncoupled mode
groups are identified. (b) Several ACs are observed as we vary
$\eta$ when modes are followed adiabatically. Inset: the spectrum
measured with a spectrometer with a higher resolution ($\sim$0.012
THz), resolving AC of two adjacent modes.} 
\label{spectrum}
\end{figure}

Let us first examine a part of spectrum obtained for $\eta$=18.7\%
as shown in Fig.\ \ref{spectrum}(a), where each peak corresponds
to a cavity mode or a quasi-eigenmode of the deformed cavity. The
spectrum in Fig.\ \ref{spectrum}(a) consists of five different
mode sequences. Modes in each sequence, marked by vertical ticks
below the spectrum, are separated by a well-defined interval
$\Delta\nu$ similar to regular modes in a symmetric cavity. This
is because all of these modes are far apart {\em accidently} in
this frequency region and thus any possible interactions among
them can be neglected. We call them {\em uncoupled}. In this
limited range of frequency we can then label these uncoupled mode
sequences  by {\em mode order} $l_0$(=1, 2, \ldots, 5) in the
increasing order of their FSRs
($\Delta\nu_1<\Delta\nu_2<\cdots<\Delta\nu_5$) in analogy to the
radial quantum number for a circular cavity \cite{Lee07PRA}.

Outside the frequency range of Fig.\ \ref{spectrum}(a), however,
some of the modes from different mode sequences would get very
close because of their different $\Delta\nu$'s and they would
interact and repel each other due to the coupling introduced by
ray chaos as to be seen later. Even in this case, we can extend
our labeling over the entire spectral range of measurement (420
THz - 530 THz) by employing the conventional assumption of
adiabatic change in $\Delta\nu$ of a given mode sequence from one
FSR to another. In order to distinguish this mode-sequence label
from the mode order defined above, we use a different notation
$l$, called {\em mode label}, such that $l$ coincide with $l_0$
only in the above limited region of frequency.

CMF and lasing spectra similar to that in Fig.\ \ref{spectrum}(a)
have been measured for various $\eta$ values from 10\% to 23\% and
all of the observed modes are labeled by the convention explained
above. A part of the results are shown in Fig.\ \ref{spectrum}(b),
where we can see some of encountering modes ($l$=2 mode and $l$=4
mode, $l$=1 mode and $l$=2 mode) undergo ACs as the cavity
deformation is varied.

In order to investigate mode dynamical properties at a {\em fixed}
deformation, we now introduce an internal parameter $n$ indexing
the recurring modes in a given mode sequence. The usefulness of
$n$ is obvious when we consider the quasi-eigenmodes of a deformed
cavity, obtained by diagonalizing a two-state effective
Hamiltonian matrix (with $\hbar$=1),
\begin{eqnarray}
H (n,\eta)= \left[\begin{array}{cc}
   \nu_p(n,\eta)-i\gamma_p (\eta)& C_{pq}(\eta) \\
  C_{pq}(\eta) & \nu_q(n,\eta)-i\gamma_q (\eta)
\end{array}\right],
\label{eqS}
\end{eqnarray}
where $\nu_{p(q)}$ and $\gamma_{p(q)}$ are frequencies and decay
rates of two {\em uncoupled} states with different mode orders,
respectively, and $C_{qp}$ is the internal coupling induced by
cavity deformation. This coupling is taken to be real because it
arises mainly from internal ray dynamics in our experiment as to
be shown later. AC along $\eta$ as shown in Fig.\ 1(b) then takes
place at $\eta_0$ satisfying $\nu_p(n,\eta_0)=\nu_q(n,\eta_0)$ for
a given $n$ if $C_{pq}>|\gamma_p-\gamma_q|/2$, a criterion for AC.
Now we consider the variation of $n$ at a given $\eta'(\ne
\eta_0)$ instead. Since states with different mode orders have
different $\Delta\nu$'s, there exists some $n_0$ satisfying
$\nu_p(n_0,\eta') \simeq \nu_q(n_0,\eta')$, for which an AC can
take place. We assume that both decay rate $\gamma_{p(q)}$ and
coupling strength $C_{pq}$ are independent of $n$ since their
dependence on frequency is not substantial in the frequency range
studied.

\begin{figure}
\includegraphics[width=3.4in]{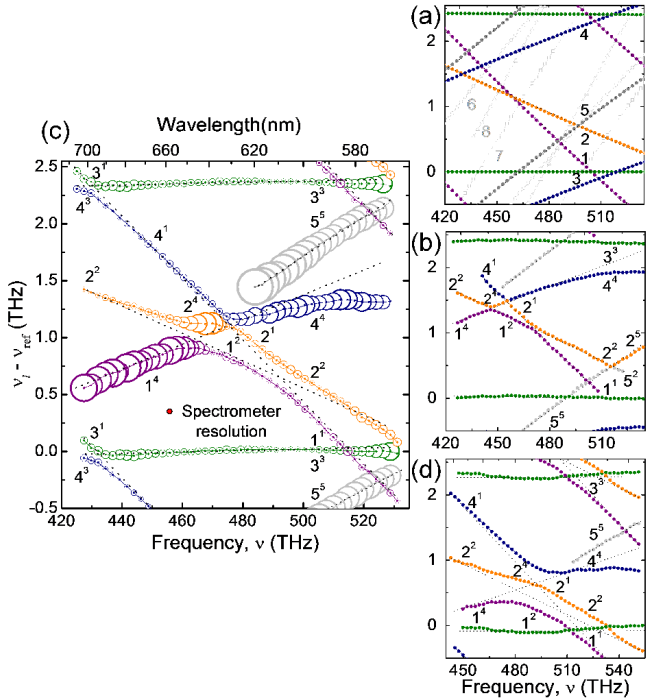}
\caption{(Color online) (a) Calculated relative frequencies of quasi-eigenmodes with radial mode order $l_0$=1, 2, $\cdots$, 8 for a circular cavity ($\eta$=0\%) with the mode frequency of $l_0$=3 as a reference. (b) Observed relative frequencies of quasi-eigenmodes for $\eta$=14.3\%. Mode frequencies more or less follow diabatic lines (dotted straight lines) except for ACs with very small splittings. We employ shorthand notation $l^{l_0}$ as explained in the text. (c) The same for $\eta$=18.7\%. More pronounced ACs with decay-rate exchange as well as ordinary crossings are observed. The diameter of the circle drawn on each data point represents the half linewidth of the corresponding mode in THz.  Red circle indicates spectrometer resolution, $\gamma_0\sim$ 0.05 THz. (d) The case of $\eta$=22.3\%. The splittings are much more larger than those of (c) and the mode frequencies deviate greatly from the diabatic lines.}
\label{avoided-summary}
\end{figure}

Mode-dynamics diagrams in Fig.\ \ref{avoided-summary} are based on
this idea of scanning $n$. We first define reference frequencies
as the resonance frequencies of $l_0$=3 whispering-gallery modes
in a circular cavity whose round trip length is the same as that
of the COM under investigation. These reference frequencies are
shown as equally-spaced vertical ticks marked as `ref' in Fig.\
\ref{spectrum}(a). We then measure the relative frequencies of the
observed quasi-eigenmodes corresponding to $n$ with respect to the
reference frequency of the same $n$ for a given $\eta$, and plot
these relative frequencies as a function of the reference
frequency corresponding to $n$. 
Mode-dynamics can be analyzed more
effectively in a mode-dynamics diagram than in Fig.\
\ref{spectrum}(b) since we can then associate the observed mode
dynamics to the relevant phase-space structure for intracavity ray
dynamics, the so-called Poinc\'{a}re surface of section (PSOS),
for a given $\eta$.

Note in Figs.\ \ref{avoided-summary}(b)-(d) that when these
quasi-eigenmodes are far apart they follow straight lines called
{\em diabatic} transition lines \cite{Takami02} even in the
presence of the internal coupling $C$ [the case of Fig.\ 1(a)]. By
shifting the internal parameter $n$, we can bring any two
quasi-eigenmodes get close and make the internal coupling come
into play. In this case, the quasi-eigenvalues deviate from the
diabatic lines significantly, exhibiting ACs.
%
Note also that the mode order $l_0$ is associated with the
uncoupled states located on the {\em diabatic} lines (straight
lines in Fig.\ 2), while the mode index $l$ is associated with
quasi-eigenmodes on {\em adiabatic} lines (exhibiting ACs in Fig.\
2). The shorthand notation $l^{l_0}$ such as $1^2$ in Fig.\ 2 is
based on this idea. Furthermore, by comparing Fig.\
\ref{avoided-summary}(a) in the case of circle with Figs.\
\ref{avoided-summary}(b)--\ref{avoided-summary}(d) for deformed
cavities, we can recognize that the modes on the $l_0$th diabatic
line must have evolved from the WGM's of radial quantum number
$l_0$ of a circular cavity. 

The diameter of the circle drawn on each data point in Fig.\
\ref{avoided-summary}(c) represents the half linewidth in THz,
directly observed with a spectrometer. It is reassuring to see
that the linewidth well before and well after an avoided crossing
is continuous along the diabatic transition line, which is a
general property of avoided crossing \cite{Takami02}. On the other
hand, in the region where avoided crossings occur, the linewidth
is an intermediate value of those well before and well after the
avoided crossing.

Another important factor to consider in Fig.\
\ref{avoided-summary} is the {\em parity} of mode. Only modes with
the same parity can interact with each other. In the frequency
range of $\nu\sim$500 THz and $0<\nu_l-\nu_{\rm ref}<$2.25 THz,
uncoupled states of $l_0$=3 and 5 have a parity different from
that of $l_0$=1, 2, and 4 states of the same $n$. This feature has
been confirmed by mode calculations by boundary element method
\cite{Wiersig03, Shim07}. This is why quasi-eigenmodes originating
from $l_0$=1, 2 and 4 states avoid each other there and why
$l_0$=1 and 2 states cross the $l_0$=3 state near ($\nu,
\nu_l-\nu_{\rm ref}$)$\sim$(510, 0)(THz) in Figs.\
\ref{avoided-summary}(b)--\ref{avoided-summary}(d). However, the
same $l_0$=1 state and another $l_0$=3 state displaced by one FSR
result in quasi-eigenstates undergoing an AC near (430, 2.3)(THz)
since any state with its mode number shifted by one ($n\rightarrow
n\pm1$) would have its parity changed to the other parity
\cite{tureci02}.

By the same reason one may expect that the same $l_0$=1 state and
another $l_0$=5 state with an one-less mode number would result in
an AC near (525, -0.25)(THz) in Figs.\ \ref{avoided-summary}(c),
(500, 0.25)(THz) in Figs.\ \ref{avoided-summary}(b), and (540,
1.4)(THz) in Figs.\ \ref{avoided-summary}(d). However, they all
exhibit a crossing instead. It is because
$C_{15}<|\gamma_5-\gamma_1|/2$, not satisfying the criterion for
AC. This example demonstrates that openness can suppress the AC in
the present internal coupling case. From this openness effect, we
can expect that the level spacing distribution would show a
delayed transition from Poisson to Wigner-like distribution in the
chaotic transition.

\begin{figure}
\includegraphics[width=3.4in]{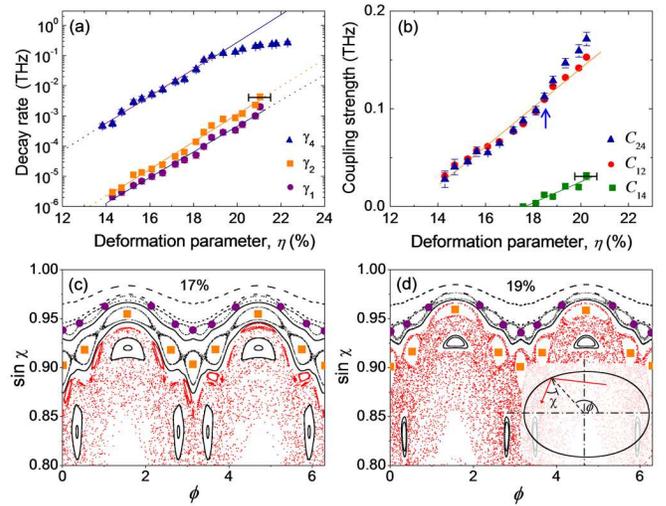}
\caption{(Color online) (a) Decay rates $\gamma_p$ of $p$=1,2,4 uncoupled states. 
(b) Coupling strength $C_{pq}$ between $p,q$(=1,2,4)
uncoupled states, restored from the observed sizes of AC and the
decay rates of modes, assuming three-mode coupling. Systematic
error in determining exact deformation is indicated with a
horizontal error bar in both (a) and (b). (c) PSOS' for $\eta$=17\% and (d) $\eta$=19\%. Large
(purple) circular and (orange) square dots represent the classical trajectories that
$l_0$=1 and 2 modes would correspond to, respectively, whereas
that of $l_0$=4 mode is embed in the chaotic sea below.
Inset: Birkhoff coordinates used in PSOS'.}
\label{PSOS}
\end{figure}

From the observed gaps of AC and the associated decay rates of
corresponding uncoupled states [Fig.\ \ref{PSOS}(a)], 
we can finally reconstruct
the {\em internal} coupling strength $C(\eta)$ between
encountering modes as shown in Fig.\ \ref{PSOS}(b). In the present case, all three $l_0$=1,2,4
modes of the same parity are coupled to each other since their
mode frequencies are not much separated in the region of
interaction. The reconstructed coupling strengths, $C_{12},
C_{24}, C_{14}$, summarized in Fig.\ \ref{PSOS}(b) are obtained by
diagonalizing a three-mode non-Hermitian symmetric Hamiltonian, a
straightforward extension of Eq.\ (1). It is the first
time to directly measure mode-mode coupling constants in an {\em
open chaotic} billiard of generic nonintegrable shape. In Fig.\
\ref{PSOS}(b) these couplings are shown to increase as the degree of
deformation increases.

\begin{figure}
\includegraphics[width=3.4in]{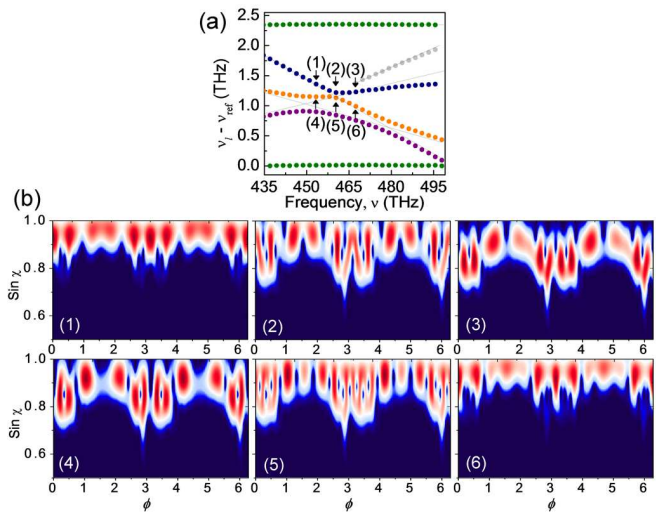}
\caption{(Color online) (a) Calculated relative frequencies of $l$=1, 2, 3, 4, 5 modes with respect to the
reference frequency. (b) Hisimi plots of the modes marked by
arrows near 460 THz where $l$=2, 4 modes undergo an AC.}
\label{husimi}
\end{figure}

Unfortunately, there is no known semiclassical theory
for enabling us to calculate the observed coupling constants. 
At best, they can be understood qualitatively in
terms of classical ray dynamics in phase space.
Following this standard practice we plot PSOS in
Figs.\ \ref{PSOS}(c) and \ref{PSOS}(d), for $\eta$=17\% and 19\%,
respectively, by using the Birkhoff coordinates with $\phi$ the
polar angle and $\chi$ the incident angle in ray tracing analysis
\cite{Lee02}. Large (purple) circular and (orange) square dots represent the classical
trajectories that $l_0$=1 and 2 modes would correspond to,
respectively, whereas that of $l_0$=4 mode is embedded in the
chaotic sea for the shown degrees of deformation. These
trajectories are inferred from phase-space distributions or Husimi
plots of the these modes \cite{Lee07PRA}. When $\eta>18\%$, as
shown in Fig.\ \ref{PSOS}(d), the classical trajectory associated
with $l_0$=2 mode no longer lie on the main integrable region,
separated from the chaotic sea by unbroken Kolmogorov-Arnold-Moser
(KAM) curve, as it did in Fig.\ \ref{PSOS}(c) for $\eta$=17\%, but
lie on islands surrounded by chaotic sea, and thus chaotic
diffusion starts to play an important role for the increased
coupling $C_{24}$ between $l_0$=2 and 4 modes as shown in Fig.\
\ref{PSOS}(a). The broken KAM curve is also responsible
for the increased coupling $C_{14}$ between $l_0$=1 and
4 modes.

The observed mode-dynamics diagrams have also been reproduced by
numerical calculations based on the boundary element method
\cite{Wiersig03, Shim07} applied for the same shape and size of
the cavity as in the experiment. The eigenvalues and associated
Husimi distributions calculated for $\eta$=0.19 are shown in Fig.\
\ref{husimi}, where we confirm that the encountering
quasi-eigenmodes exchange their mode distributions upon avoided
crossing [(1)$\leftrightarrow$(6) and  (4)$\leftrightarrow$(3)]
and at the closest encounter the resulting modes [(2) and (5)] are
linear superpositions of the modes well before and well after the
avoided crossing, thus leading to delocalized eigenfunctions
\cite{Takami02, Noid83}.

In conclusion, we have developed an spectroscopic
technique to enable experimental investigation of mode-dynamics
evolution along the chaotic transition in open chaotic billiards.
The observed mode-dynamics evolution shows that openness tends to
suppress avoided crossings compared to the closed billiard cases.
We could directly measure the coupling strengths
induced by ray chaos among encountering modes. Our
measurements would serve as a valuable asset for anticipated but
currently-nonexisting semiclassical theory for coupling  strengths
between modes in a mixed phase space.

This work was supported by National Research Laboratory Grant and
by WCU Grant. S.W.K. was supported by KRF Grant
(2006-005-J02804) and by KOSEF Grant (R01-2005-000-10678-0).
S.Y.L. was supported by BK21 program.



\begin{thebibliography}{99}

\bibitem{Q-chaos}
H.-J.\ St\"{o}ckmann,  {\it Quantum Chaos: an Introduction} (Cambridge Univ.\ Press, Cambridge, 1999).

\bibitem{stoeckmann}
H.-J.\ St\"{o}ckmann and J.\ Stein, Phys.\ Rev.\ Lett.\ {\bf 64}, 2215 (1990); H.-D.\ Gr\"{a}f {\it et al.}, Phys.\ Rev.\ Lett.\ {\bf 69}, 1296 (1992).

\bibitem{backer}
A.\ B\"{a}cker {\it et al.}, Phys.\ Rev.\ Lett.\ {\bf 100}, 174103 (2008).

\bibitem{Persson}
E.\ Persson, I.\ Rotter, H.\ -J.\ St\"{o}ckmann, and M.\ Barth, Phys.\ Rev.\ Lett.\ {\bf 85}, 2478 (2000).

\bibitem{Dembowiski}
C.\ Dembowski {\it et al.}, Phys.\ Rev.\ Lett.\ {\bf 86}, 787 (2001).

\bibitem{noeckel}
J.\ U.\ N\"{o}ckel, A.\ D.\ Stone, G.\ Chen, H.\ L.\ Grossman, and R.\ K.\ Chang, Opt.\ Lett.\ {\bf 21}, 1609 (1996).

\bibitem{Yang-RSI06}
J.\ Yang {\it et al.}, Rev.\ Sci.\ Instrum.\ {\bf77}, 083103 (2006).

\bibitem{Moon-OE08} S.\ Moon {\it et al.}, Optics Express {\bf 16}, 11007 (2008).

\bibitem{Lee02}
S.-B.\ Lee, J.-H.\ Lee, J.-S.\ Chang, H.-J.\ Moon, S.\ W.\ Kim, K.\ An, Phys.\ Rev.\ Lett.\ {\bf 88}, 033903 (2002).

\bibitem{Lee07PRA}
S.-B.\ Lee {\it et al.}, Phys.\ Rev.\ A {\bf 75}, 011802(R) (2007).

\bibitem{Takami02}
T.\ Takami, Phys.\ Rev.\ Lett.\ {\bf 68}, 3371 (1992).

\bibitem{Wiersig03} J.\ Wiersig, J.\ Opt.\ A {\bf 5}, 53 (2003).

\bibitem{Shim07}
J.-B.\ Shim {\it et al.}, J.\ Phys.\ Soc.\ Jpn.\ {\bf 76}, 114005 (2007).

\bibitem{tureci02}
H.\ E.\ Tureci, H.\ G.\ L.\ Schwefel, A.\ D.\ Stone, and E.\ E.\ Narimanov, Opt.\ Exp.\ {\bf 10}, 752 (2002).

\bibitem{Noid83}
D.\ W.\ Noid, M.\ L.\ Koszykowski, and R.\ A.\ Marcus, J.\ Chem.\ Phys.\ {\bf 78}, 4018 (1983).

\end{thebibliography}
\end{document}